\documentclass[10pt]{article}         
\usepackage{amsmath,amsfonts,amssymb}   

\title{Virial Theorems and Virial Stresses of Micropolar Media}
\author{Martin Ostoja-Starzewski}

\begin{document}

\maketitle

\begin{center}
\small Department of Mechanical Science \& Engineering\\
\small also Institute for Condensed Matter Theory and Beckman Institute\\
\small University of Illinois at Urbana-Champaign\\
\small Urbana, IL 61801, U.S.A.\\
\small e-mail: martinos@uiuc.edu
\end{center}

\begin{abstract}
A generalization of virial theorems and virial stresses to micropolar
continuum mechanics is explored. \ The linear momentum balance in dyadic
product with translation leads to (i) the \textit{first} \textit{virial
theorem of micropolar\ continuum mechanics} involving the infinite-time
limit of the kinetic translational energy and (ii)\ a classical formula for
computing the virial force-stress known in molecular dynamics. \ The angular
momentum balance in dyadic product with rotation leads to (i) the \textit{%
second} \textit{virial theorem of micropolar\ continuum mechanics} involving
the infinite-time limit of the kinetic rotational energy and (ii)\ a \textit{%
virial couple-stress} along with a formula for its computation. \ The latter
stress is also uncovered in the dyadic product of linear momentum balance
with rotation. \ The virial force-stress and virial couple-stress contain,
respectively, the Reynolds force-stress and turbulent couple-stress. 
\end{abstract}

\section{Introduction}

This communication reports a generalization of the virial theorem and virial
stress from classical to micropolar continuum mechanics. \ The study is set
in the spatial (Eulerian)\ description [Eringen, 2001; \L ukaszewicz, 1999]
in $\mathbb{E}^{d}$, $d=2$ or $3$. \ For reference, the kinematics of such a
continuum is described through a field with two degrees of freedom $\left(
u_{i},\varphi _{i}\right) $ and their rates:%
\begin{equation}
\text{velocity: \ }\dot{u}_{i}\text{; \ \ \ \ \ microrotation: \ }\dot{%
\varphi}_{i}.  \label{degrees of freedom}
\end{equation}%
The overdot stands for the material time derivative. \ Standard index as
well as symbolic tensor notations are used. \ 

The local form of linear momentum balance reads 
\begin{equation}
\rho \dot{v}_{i}=\rho f_{i}+\tau _{ki},_{k},  \label{local form linear}
\end{equation}%
where $\rho $ is the mass density, $f_{i}$ is the body force per unit mass,
and $\tau _{ji}$\ is the (generally non-symmetric) Cauchy force-stress. \
The latter is related to the force-traction on any surface element specified
by the outer normal $n_{k}$\ through $t_{i}^{\left( \mathbf{n}\right) }=\tau
_{ji}n_{j}$.

The local form of angular momentum balance, after accounting for (\ref{local
form linear}), reads 
\begin{equation}
\rho J\frac{D\dot{\varphi}_{i}}{Dt}=\rho g_{i}+\mu _{ki},_{k}+T_{i}\text{ \
\ where \ \ }T_{i}=\epsilon _{ijk}\tau _{jk},  \label{local form angular}
\end{equation}%
where $J$ is the (micro)inertia tensor of a particle, $g_{i}$ is the body
torque per unit mass, and\ $\mu _{ki}$ is the couple-stress, which is
related to the surface moment-traction $m_{i}^{\left( \mathbf{n}\right) }$
through $m_{i}^{\left( \mathbf{n}\right) }=\mu _{ki}n_{k}$. \ If $T_{i}=0$, $%
\tau _{jk}$\ becomes symmetric, in which case we write $\sigma _{jk}$\
instead. \ In (\ref{local form linear}) and henceforth, we work with a 
\textit{microisotropic} medium, i.e. where $J_{ij}=J\delta _{ij}$.

In general, we consider the material properties to be constant, while the
displacement, rotation, force-stress, and couple stress, as well as their
derivatives, to be tensor-valued random fields [Malyarenko and
Ostoja-Starzewski, 2019]. \ Denoting by $F(\mathbf{x},t,\omega )$\ a
realization of any such random field, it is clearly seen to be parametrized
by the (spatial) position $\mathbf{x}$ and time $t$, where $\omega $\ is a
realization drawn from a probability space $\left( \Omega ,\digamma ,\mathsf{%
P}\right) $. \ Thus, there are three basic types of averaging:

(i)\ \textit{Space (volume)\ averaging} of volume integrals:%
\begin{equation}
\overline{F}_{V}(\mathbf{x},t,\omega ):=\frac{1}{V}\int_{\mathcal{B}}F(%
\mathbf{x}^{\prime },t,\omega )dV,
\end{equation}%
where $V=|\mathcal{B}|$ is the volume of domain $\mathcal{B}$ having a
bounding surface $\partial \mathcal{B}$. \ In the language of stochastic
processes, this is \textit{local averaging} transforming the original random
field $F$ into a new, smoother random field $\overline{F}_{V}$. \ In
general, $\overline{F}_{V}$\ is a function of $\mathbf{x}$ but, if spatial
ergodicity (i.e., equivalence of spatial averaging with time averaging
and/or ensemble averaging) is assumed, this dependence vanishes. \ This
situation corresponds to passing from a statistical to a representative
volume element (RVE), with the RVE size heavily dependent on the random
microstructural properties involved [Ostoja-Starzewski \textit{et al}.,
2016]. \ In the following, to simplify the notation, we will simply write $%
\overline{F}$.

(ii)\ \textit{Time averaging,} wherein fluctuations observed on a fine time
scale get smoothed (smeared) on a coarse time scale;%
\begin{equation}
F_{T}(\mathbf{x},t,\omega ):=\frac{1}{T}\int_{0}^{T}F(\mathbf{x},t^{\prime
},\omega )dV.
\end{equation}%
In the language of stochastic processes, this is \textit{local averaging} in
the time domain, transforming the original random field $F$ into a new,
smoother-in-time random field $F_{T}$. \ In general, $F_{T}$\ is still a
random function of $t$ but, if temporal ergodicity (i.e., equivalence with
space and/or ensemble averaging) is assumed, this dependence vanishes. \ The
latter situation typically implies taking the $T\rightarrow \infty $\ limit.

(iii)\ \textit{Ensemble (statistical) averaging}.

While in either case, one obtains a coarser description than the original
one, in the present study only space averaging over arbitrary volumes and
time averaging over infinitely long times is considered. \ The latter
operation, called \textit{infinite-time averaging}, is indicated by%
\begin{equation}
\left\{ F(\mathbf{x},t,\omega )\right\} _{\infty }=\lim_{T\rightarrow \infty
}\left\{ F(\mathbf{x},t,\omega )\right\} _{T}.
\label{infinite time averaging}
\end{equation}%
The reason for considering such a limit as well as the spatial averaging
resides in our focus on the exploration of the virial theorem(s) and virial
stress(es) in micropolar media. \ The ensuing study aims to generalize the
deterministic exposition of those topics in the setting of classical media
[Podio-Guidugli, 2019], whereby we follow a similar line of reasoning. \ The
statistical exposition in the vein of that paper is not discussed here.

\bigskip

\section{Consequences of linear momentum balance}

First, we take a dyadic product of a test function $\phi _{j}$ with (\ref%
{local form linear}) and integrate over an arbitrary (and simply-connected)
domain $\mathcal{B}$\ to get%
\begin{equation}
\int_{\mathcal{B}}\left( \phi _{j}\rho f_{i}+\phi _{j}\tau _{ki,k}\right)
dV=\int_{\mathcal{B}}\phi _{j}\rho \dot{v}_{i}dV.  \label{linear integrated}
\end{equation}%
We now consider two choices for the test function.

\subsection{Translation as test function}

Taking translation as the test function ($\phi _{j}=x_{j}$) in (\ref{linear
integrated}), we perform these steps%
\begin{equation}
\begin{array}{c}
\int_{\mathcal{B}}\left( x_{j}\rho f_{i}+x_{j}\tau _{ki,k}\right) dV=\int_{%
\mathcal{B}}\left[ \rho \left( x_{j}v_{i}\right) ^{\cdot }-\rho \left( \dot{x%
}_{j}v_{i}\right) \right] dV \\ 
\int_{\mathcal{B}}x_{j}\rho f_{i}dV+\int_{\mathcal{B}}\left[ \left(
x_{j}\tau _{ki}\right) ,_{k}-x_{j},_{k}\tau _{ki}\right] dV=\frac{d}{dt}%
\left( \int_{\mathcal{B}}x_{j}\rho v_{i}dV\right) -\int_{\mathcal{B}}\left( 
\dot{x}_{j}\rho v_{i}\right) dV \\ 
\int_{\mathcal{B}}x_{j}\rho f_{i}dV+\int_{\partial \mathcal{B}}x_{j}\tau
_{ki}n_{k}dS-\int_{\mathcal{B}}\tau _{ji}dV=\frac{d}{dt}\left( \int_{%
\mathcal{B}}x_{j}\rho v_{i}dV\right) -V\overline{\rho v_{j}v_{i}} \\ 
\int_{\mathcal{B}}x_{j}\rho f_{i}dV+\int_{\partial \mathcal{B}}x_{j}\tau
_{ki}n_{k}dS-\int_{\mathcal{B}}\tau _{ji}dV=\frac{d}{dt}\left( \int_{%
\mathcal{B}}x_{j}\rho v_{i}dV\right) -V\overline{\rho v_{j}v_{i}}.%
\end{array}%
\end{equation}%
\ Hence, we can write%
\begin{equation}
\int_{\mathcal{B}}\rho x_{j}f_{i}dV+\int_{\partial \mathcal{B}%
}x_{j}t_{i}^{\left( \mathbf{n}\right) }dS-V\tau _{ji}^{\ast }=\frac{d}{dt}%
\left( \int_{\mathcal{B}}\rho x_{j}v_{i}dV\right) ,  \label{dyadic of linear}
\end{equation}%
where 
\begin{equation}
\tau _{ji}^{\ast }:=\overline{\tau _{ji}}-\overline{\rho v_{j}v_{i}}
\label{virial force-stress}
\end{equation}%
is the \textit{virial stress}. \ The term $\overline{\rho v_{j}v_{i}}$\
contains $\overline{\rho v_{j}^{\prime }v_{i}^{\prime }}$, as can be seen
from the Reynolds decomposition ($v_{j}=\overline{v_{j}}+v_{j}^{\prime }$
with $\overline{v_{j}^{\prime }}=0$). \ Depending on a convention, $%
\overline{\rho v_{j}v_{i}}$ or $\overline{\rho v_{j}^{\prime }v_{i}^{\prime }%
}$ (with a plus or minus sign)\ is called the Reynolds stress. \ From the
standpoint of turbulent flows, $\tau _{ji}^{\ast }$\ is the \textit{modified
stress} in the sense that the linear momentum, angular momentum, and energy
balances are unchanged in form upon volume averaging [Ostoja-Starzewski,
2021]. \ 

Going from continuum to discrete systems such as molecular dynamics, (\ref%
{virial force-stress})\ is consistent with the very well-known prescription
for computation of the instantaneous volume-averaged virial (generally
non-symmetric)\ force-stress%
\begin{equation}
\tau _{ji}^{\ast }:=\frac{1}{V}\sum_{k\in \mathcal{B}}\left[ \frac{1}{2}%
\sum_{l\in \mathcal{B}}\left( x_{i}^{\left( l\right) }-x_{i}^{\left(
k\right) }\right) f_{j}^{\left( kl\right) }-m^{\left( k\right) }\left(
v_{j}^{\left( k\right) }-\overline{v_{j}}\right) \left( v_{i}^{\left(
k\right) }-\overline{v_{i}}\right) \right] ,
\label{discrete virial force-stress}
\end{equation}%
where

$k$ and $l$ are particles in the domain $\mathcal{B}$,

$m^{\left( k\right) }$ is the mass of particle $k$,

$v_{j}^{\left( k\right) }$ is the $j$-th component of the velocity of
particle $k$,

$\overline{v_{j}}$ is the $j$-th component of the average velocity of
particles in the volume,

$x_{i}^{\left( k\right) }$ is the $i$-th component of the position of
particle $k$, and

$f_{j}^{\left( kl\right) }$ is the $j$-th component of the force applied on
particle $k$ by particle $l$.

\bigskip

No symmetry of the Cauchy stress is required in the above derivation. \
Also, (\ref{virial force-stress}) does not imply the symmetry of $\tau
_{ji}^{\ast }$\ unless all the body torques and couple-stresses are zero:
this is seen by applying (\ref{local form angular}) on the lengthscale $L=%
\sqrt[d]{V}$ ($d=2$ or $3$)\ on which the space averaging is carried out. \
This would imply writing $\rho J_{ij}\frac{D\dot{\varphi}_{j}}{Dt}=\rho
g_{i}+\mu _{ki,k}^{\ast }+\epsilon _{ijk}\tau _{jk}^{\ast }$, where $\mu
_{ki}^{\ast }$\ is given by (\ref{virial couple-stress}) below. \ In
essence, the formula (2.5) alone is not sufficient to judge whether this
force-stress will be symmetric or not.

Taking the trace of (\ref{dyadic of linear})\ we get:%
\begin{equation}
\int_{\mathcal{B}}\rho x_{i}f_{i}dV+\int_{\partial \mathcal{B}%
}x_{i}t_{i}^{\left( \mathbf{n}\right) }dS-\int_{\mathcal{B}}\tau _{ii}dV=%
\dot{W}_{u}\left( \mathcal{B}\right) -2K_{u}\left( \mathcal{B}\right) ,
\end{equation}%
where%
\begin{equation}
W_{u}\left( \mathcal{B}\right) :=V\text{ }\overline{\rho x_{i}v_{i}},\text{
\ \ }K_{u}\left( \mathcal{B}\right) :=\frac{V}{2}\overline{\rho v_{i}v_{i}},
\label{kinetic translational energy}
\end{equation}%
with the latter being the \textit{kinetic translational energy}.

Performing the infinite-time averaging (\ref{infinite time averaging}), with
the argument given in [Podio-Guidugli, 2019], the above becomes%
\begin{equation}
\left\{ \int_{\mathcal{B}}\rho x_{i}f_{i}dV+\int_{\partial \mathcal{B}%
}x_{i}t_{i}^{\left( \mathbf{n}\right) }dS-\int_{\mathcal{B}}\tau
_{ii}dV\right\} _{\infty }=-2\left\{ K_{u}\left( \mathcal{B}\right) \right\}
_{\infty },  \label{first virial theorem in micropolar}
\end{equation}%
which, in view of Section 3 below, is called the \textit{first} \textit{%
virial theorem of micropolar\ continuum mechanics}. \ The arguments
accompanying this infinite time limit include the assumption of kinematic
uniform or periodic boundary conditions (Costanzo \textit{et al}., 2005).

Taking the dual vector of (\ref{dyadic of linear})\ we find two relations to
be satisfied by the skew-symmetric part of the Cauchy stress tensor and,
essentially, linking all the quantities appearing in (\ref{local form linear}%
) and (\ref{local form angular}):%
\begin{equation}
\begin{array}{c}
\int_{\mathcal{B}}\left( \rho J\ddot{\varphi}_{k}-\rho g_{k}-\mu
_{ik},_{i}\right) dV=\int_{\mathcal{B}}T_{k}dV \\ 
=\int_{\mathcal{B}}\rho \epsilon _{kji}x_{j}f_{i}dV+\int_{\partial \mathcal{B%
}}\epsilon _{kji}x_{j}t_{i}^{\left( \mathbf{n}\right) }dS-\int_{\mathcal{B}%
}\rho \epsilon _{kji}x_{j}\dot{v}_{i}dV.%
\end{array}%
\end{equation}

In the special case of micropolar effects in (\ref{local form linear})-(\ref%
{local form angular}) being absent, $W_{\varphi }\left( \mathcal{B}\right)
=0 $ and $K_{\varphi }\left( \mathcal{B}\right) =0$ in (\ref{kinetic
rotational energy}), so that (\ref{first virial theorem in micropolar})
simplifies to the virial theorem of classical continuum mechanics%
\begin{equation}
\left\{ \int_{\mathcal{B}}\rho x_{i}f_{i}dV+\int_{\partial \mathcal{B}%
}x_{i}t_{i}^{\left( \mathbf{n}\right) }dS-\int_{\mathcal{B}}\sigma
_{ii}dV\right\} _{\infty }=-2\left\{ K\left( \mathcal{B}\right) \right\}
_{\infty }.
\end{equation}%
\textit{Note}: there are various ways for couple-stress effects to arise,
typically involving some internal material structure, and these include:

- presence of moment interactions in granular media or suspensions, e.g.
[Cowin, 1974; Eringen, 2001; Vardoulakis, 2019];

- presence of defects, phase boundaries, e.g. [Chen \textit{et al}., 2006;
Rigelesaiyin \textit{et al}. 2018];

- homogenization of random media, e.g. [Trovalusci \textit{et al.}, 2015];

- fractal media in which the fractal structure (homogenized in the vein of
dimensional regularization) is anisotropic [Li \&\ Ostoja-Starzewski, 2011].

The symmetric part of (\ref{dyadic of linear})\ was discussed in
[Podio-Guidugli, 2019].

\subsection{Rotation as test function}

Taking rotation as the test function ($\phi _{j}=x_{j}$) in (\ref{linear
integrated}), we obtain%
\begin{equation}
\int_{\mathcal{B}}\varphi _{j}\rho f_{i}dV+\int_{\partial \mathcal{B}%
}\varphi _{j}t_{i}^{\left( \mathbf{n}\right) }dS-\int_{\mathcal{B}}\varphi
_{j},_{k}\tau _{ki}dV+\int_{\mathcal{B}}\rho \text{ }\dot{\varphi}%
_{j}v_{i}dV=\frac{d}{dt}\left( \int_{\mathcal{B}}\rho \varphi
_{j}v_{i}dV\right)
\end{equation}%
Multiplying this equation by $J$ yields%
\begin{equation}
\int_{\mathcal{B}}\rho J\varphi _{j}f_{i}dV+\int_{\partial \mathcal{B}%
}J\varphi _{j}t_{i}^{\left( \mathbf{n}\right) }dS-\int_{\mathcal{B}}J\varphi
_{j},_{k}\tau _{ki}dV-\left( -\int_{\mathcal{B}}\rho J\dot{\varphi}%
_{j}v_{i}dV\right) =\frac{d}{dt}\left( \int_{\mathcal{B}}\rho J\varphi
_{j}v_{i}dV\right) .  \label{rotation-linear}
\end{equation}%
The new term in parenthesis on the left-hand side, $-V\overline{\rho J\dot{%
\varphi}_{j}v_{i}}$, will independently arise in Section 3.2.

\section{Consequences of angular momentum balance}

Consider a dyadic product of test function $\phi _{j}$ with the vector (\ref%
{local form angular}) and integrate over $\mathcal{B}$ to get%
\begin{equation}
\int_{\mathcal{B}}\phi _{j}\left( \rho g_{i}+\mu _{ki},_{k}+\epsilon
_{ink}\tau _{nk}\right) dV=\int_{\mathcal{B}}\phi _{j}\rho J\ddot{\varphi}%
_{i}dV.  \label{angular integrated}
\end{equation}%
Again, there are two basic options for choosing the test function.

\subsection{Rotation as test function}

On account of the Green-Gauss Theorem and taking rotation as the test
function ($\phi _{j}=\varphi _{j}$),\ we find%
\begin{equation}
\begin{array}{c}
\int_{\mathcal{B}}\varphi _{j}\rho g_{i}dV+\int_{\partial \mathcal{B}%
}\varphi _{j}m_{i}^{\left( \mathbf{n}\right) }dS-\int_{\mathcal{B}}\varphi
_{j},_{k}\mu _{ki}dV+\int_{\mathcal{B}}\varphi _{j}\epsilon _{ink}\tau
_{nk}dV \\ 
=\frac{d}{dt}\left( \int_{\mathcal{B}}\rho J\varphi _{j}\dot{\varphi}%
_{i}dV\right) -\int_{\mathcal{B}}\rho J\dot{\varphi}_{j}\dot{\varphi}_{i}dV.%
\end{array}
\label{dyadic of angular}
\end{equation}%
Taking the trace of (\ref{dyadic of angular})\ we obtain (with $T_{i}$\
defined in (\ref{local form angular})$_{2}$)%
\begin{equation}
\int_{\mathcal{B}}\varphi _{i}\rho g_{i}dV+\int_{\partial \mathcal{B}%
}\varphi _{i}m_{i}^{\left( \mathbf{n}\right) }dS-\int_{\mathcal{B}}\varphi
_{i},_{k}\mu _{ki}dV+\int_{\mathcal{B}}\varphi _{i}T_{i}dV=\dot{W}_{\varphi
}\left( \mathcal{B}\right) -2K_{\varphi }\left( \mathcal{B}\right) ,
\end{equation}%
where 
\begin{equation}
W_{\varphi }\left( \mathcal{B}\right) :=V\text{ }\overline{\rho J\varphi _{j}%
\dot{\varphi}_{i}},\text{ \ \ }K_{\varphi }\left( \mathcal{B}\right) :=\frac{%
V}{2}\overline{\rho J\dot{\varphi}_{i}\dot{\varphi}_{i}},
\label{kinetic rotational energy}
\end{equation}%
with the latter being the \textit{kinetic rotational energy}; compare with (%
\ref{kinetic translational energy}).

Under the infinite-time averaging (\ref{infinite time averaging}), we find%
\begin{equation}
\left\{ \int_{\mathcal{B}}\rho \varphi _{i}g_{i}dV+\int_{\partial \mathcal{B}%
}\varphi _{i}m_{i}^{\left( \mathbf{n}\right) }dS-\int_{\mathcal{B}}\varphi
_{i},_{k}\mu _{ki}dV+\int_{\mathcal{B}}\varphi _{i}T_{i}dV\right\} _{\infty
}=-2\left\{ K_{\varphi }\left( \mathcal{B}\right) \right\} _{\infty },
\label{second virial theorem in micropolar}
\end{equation}%
which we call the \textit{second} \textit{virial theorem of micropolar\
continuum mechanics}.

\bigskip

\subsection{Translation as test function}

Considering $\phi _{j}=x_{j}$\ in (\ref{angular integrated})\ we get:%
\begin{equation}
\begin{array}{c}
\int_{\mathcal{B}}x_{j}\rho g_{i}dV+\int_{\partial \mathcal{B}%
}x_{j}m_{i}^{\left( \mathbf{n}\right) }dS-\int_{\mathcal{B}}\mu
_{ji}dV+\int_{\mathcal{B}}x_{j}\epsilon _{ink}\tau _{nk}dV= \\ 
\frac{d}{dt}\left( \int_{\mathcal{B}}\rho Jx_{j}\dot{\varphi}_{i}dV\right)
-\int_{\mathcal{B}}\rho J\dot{x}_{j}\dot{\varphi}_{i}dV,%
\end{array}%
\end{equation}%
which can be written as%
\begin{equation}
\int_{\mathcal{B}}x_{j}\rho g_{i}dV+\int_{\partial \mathcal{B}%
}x_{j}m_{i}^{\left( \mathbf{n}\right) }dS-V\mu _{ji}^{\ast }+\int_{\mathcal{B%
}}x_{j}T_{i}dV=\frac{d}{dt}\left( \int_{\mathcal{B}}\rho Jx_{j}\dot{\varphi}%
_{i}dV\right) ,  \label{translation-angular}
\end{equation}%
where 
\begin{equation}
\mu _{ji}^{\ast }:=\overline{\mu _{ji}}-\overline{\rho J\dot{\varphi}%
_{j}v_{i}}  \label{virial couple-stress}
\end{equation}%
is the \textit{virial couple-stress}. \ As can be seen from the Reynolds
decomposition of $\varphi _{j}$ and $v_{i}$, the term $-\overline{\rho J\dot{%
\varphi}_{j}v_{i}}$\ contains 
\begin{equation}
S_{ji}:=-\overline{\rho J\dot{\varphi}_{j}^{\prime }v_{i}^{\prime }}.
\end{equation}%
which was called the \textit{turbulent couple-stress} in\
[Ostoja--Starzewski, 2021]. \ No microisotropy restriction was assumed
there, so the formula for $S_{ji}$ was a bit more general. \ This
couple-stress may be thought of as an analogue of the Reynolds stress $-%
\overline{\rho v_{j}^{\prime }v_{i}^{\prime }}$\ brought up in Section 2.1.
\ 

Interestingly, $-\overline{\rho J\dot{\varphi}_{j}v_{i}}$\ appeared via a
completely different route in (\ref{rotation-linear}). \ As discussed in
[Ostoja--Starzewski, 2021], from the standpoint of turbulent flows, $\mu
_{ji}^{\ast }$\ is the modified stress in the sense that, if it replaces $%
\mu _{ji}$\ along with the heat flux and internal energy density\ being also
replaced by modified expressions, then the angular momentum and energy
balance equations are unchanged in form upon volume averaging. \ 

Again, going from continuum to discrete systems such as the
molecular/particle dynamics, (\ref{virial couple-stress})\ gives a
prescription for computation of the instantaneous volume-averaged virial
couple-stress%
\begin{equation}
\mu _{ji}^{\ast }:=\frac{1}{V}\sum_{k\in \mathcal{B}}\left[ \frac{1}{2}%
\sum_{l\in \mathcal{B}}\left( x_{i}^{\left( l\right) }-x_{i}^{\left(
k\right) }\right) m_{j}^{\left( kl\right) }-J^{\left( k\right) }\left( \dot{%
\varphi}_{j}^{\left( k\right) }-\overline{\dot{\varphi}_{j}}\right) \left(
v_{i}^{\left( k\right) }-\overline{v_{i}}\right) \right] ,
\label{discrete virial couple-stress}
\end{equation}%
where

$k$ and $l$ are particles in the domain $\mathcal{B}$,

$J^{\left( k\right) }$ is the mass moment of inertia of particle $k$,

$v_{j}^{\left( k\right) }$ is the $j$-th component of the velocity of
particle $k$,

$\overline{v_{j}}$ is the $j$-th component of the average velocity of
particles in the volume,

$\dot{\varphi}_{i}^{\left( k\right) }$ is the $i$-th component of the
velocity of particle $k$,

$\overline{\dot{\varphi}_{i}}$ is the $j$-th component of the average
velocity of particles in the volume,

$x_{i}^{\left( k\right) }$ is the $i$-th component of the position of
particle $k$, and

$m_{j}^{\left( kl\right) }$ is the $j$-th component of the moment applied on
particle $k$ by particle $l$.

\bigskip

Note the analogy between the formulas (\ref{discrete virial couple-stress})
and (\ref{discrete virial force-stress}).

\bigskip

\section{Conclusions}

This communication generalizes the virial theorem and virial stress to
micropolar continuum mechanics. \ The study hinges on exploring the dyadic
products of linear and angular momentum balances with translational and
rotational test functions. \ In generic terms, the results are summarized as
follows.

\begin{itemize}
\item The linear momentum balance with translation leads to:

(i)\ A virial theorem known in classical continuum theories involving the
infinite-time limit of the kinetic translational energy. \ It is called the 
\textit{first} \textit{virial theorem of micropolar\ continuum mechanics}. \ 

(ii)\ A classical formula for computing the virial force-stress known in
molecular dynamics.

\item The angular momentum balance with rotation leads to a new virial
theorem involving the infinite-time limit of the kinetic rotational energy.
\ It is called the \textit{second} \textit{virial theorem of micropolar\
continuum mechanics}.

\item The angular momentum balance with translation leads to a \textit{%
virial couple-stress} along with a formula for its computation. \ The same
stress arises independently in the linear momentum balance with rotation.

\item The virial couple-stress contains the \textit{turbulent couple-stress}%
, just like the virial force-stress contains the Reynolds stress.

\item The virial force-stress and couple-stress formulas do not require
`infinitely' large volumes (on the RVE level) and can be used to assess
tensor-valued random fields on mesoscale levels as in [Ostoja-Starzewski and
Laudani [2020].
\end{itemize}

These results have applicability in computational studies as well as a
stepping-stone to micropolar extensions of various complex phenomena such as
presented in, say, [Ganghoffer, 2015; Chen and Diaz, 2016, 2018].

\bigskip

{\Large Acknowledgment}

Comments by two anonymous reviewers are gratefully acknowledged.

\bigskip

{\Large References}

Chen, Y., Lee, J. and Xiong, L. [2006] "Stresses and strains at nano-/micro
scales," \textit{J. Mech. Mater. Struct.} \textbf{1}(4), 705-723.

Chen, Y. and Diaz, A. [2016] "Local momentum and heat fluxes in transient
transport processes and inhomogeneous systems," \textit{Phys. Rev. E} 
\textbf{94}(5), 053309.

Chen, Y. and Diaz, A. [2018] "Physical foundation and consistent formulation
of atomiclevel fluxes in transport processes," \textit{Phys. Rev. E} \textbf{%
98}(5), 052113.

Costanzo, F., Gray, G.L. and Andia, P.C. [2005] "On the definitions of
effective stress and deformation gradient for use in MD:\ Hill's
macro-homogeneity and the virial theorem,"\ \textit{International Journal of
Engineering Science} \textbf{43}, 533--555.

Cowin, S.C. [1974] "The theory of polar fluids," \textit{Advances in Applied
Mechanics} \textbf{14}, 279-347.

Eringen, A.C. [2001] \textit{Microcontinuum Field Theories II:\ Fluent Media}%
. Springer-Verlag, New York.

Ganghoffer, J.-F. [2015] "Spatial and material stress tensors in continuum
mechanics of growing solid bodies,"\ \textit{MEMOCS} \textbf{3}(4), 341-363.

Li, J. and Ostoja-Starzewski, M. [2011] "Micropolar continuum mechanics of
fractal media,"\ \textit{Int. J. Eng. Sci.} (A.C. Eringen special issue) 
\textbf{49}, 1302-1310.

\L ukaszewicz, G. [1999] \textit{Micropolar Fluids: Theory and Applications}%
. Birkh\"{a}user, Basel.

Malyarenko, A. and Ostoja-Starzewski, M. [2019] \textit{Tensor-Valued Random
Fields for Continuum Physics}. Cambridge University Press, Cambridge.

Ostoja-Starzewski, M., Kale, S., Karimi, P., Malyarenko, A., Raghavan, B.,
Ranganathan, S.I. and J. Zhang, J. [2016] "Scaling to RVE in random media,"\ 
\textit{Advances in Applied Mechanics} \textbf{49} (eds. S.P.A. Bordas and
D.S. Balint), 111-211.

Ostoja-Starzewski, M. and Laudani, R. [2020] "Violations of the
Clausius-Duhem inequality in Couette flows of granular media," \textit{Proc.
R. Soc. A} \textbf{476}: 20200207.

Ostoja-Starzewski, M. [2021] "Averaging of turbulent micropolar media:
Turbulent couple-stress, heat flux, and energy,"\ \textit{ZAMP} \textbf{72}%
:106.

Podio-Guidugli, P. [2018] "The virial theorem:\ A pocket primer,"\ \textit{%
J. Elast.} \textbf{137}, 219-235.

Rigelesaiyin, J., Diaz, A., Li, W., Xiong, L., Chen, Y. [2018] "Asymmetry of
the atomic-level stress tensor in homogeneous and inhomogeneous materials," 
\textit{Proc. R. Soc. A} \textbf{474}: 20180155.

Trovalusci, P., Ostoja-Starzewski, M., De Bellis, M.L., Murrali, A. [2015]
"Scale-dependent homogenization of random composites as micropolar
continua,"\ \textit{Europ. J. Mech.-A/Solids} \textbf{49}, 396-407.

Vardoulakis, I. [2019] \textit{Cosserat Continuum Mechanics: with
Applications to Granular Media} (Springer Science \& Business Media, Basel).

\end{document}